\documentclass{PoS}

\title{Front-end electronics for the GAPS tracker}

\ShortTitle{FE for the GAPS tracker}

\author{\speaker{Valentina Scotti}\\
		Universit\`{a} degli Studi di Napoli Federico II, Dipartimento di Fisica, Italy\\
        Istituto Nazionale di Fisica Nucleare - Sezione di Napoli, Italy\\
        E-mail: \email{scottiv@na.infn.it}}

\author{Alfonso Boiano\\
        Istituto Nazionale di Fisica Nucleare - Sezione di Napoli, Italy}

\author{Lorenzo Fabris \\ Oak Ridge National Laboratory, TN, USA}

\author{Massimo Manghisoni \\ University of Bergamo, Italy\\ Istituto Nazionale di Fisica Nucleare - Sezione di Pavia, Italy}

\author{Giuseppe Osteria\\         Istituto Nazionale di Fisica Nucleare - Sezione di Napoli, Italy}

\author{Francesco Perfetto\\         Istituto Nazionale di Fisica Nucleare - Sezione di Napoli, Italy}

\author{Valerio Re\\ University of Bergamo, Italy\\ Istituto Nazionale di Fisica Nucleare - Sezione di Pavia, Italy}
\author{Elisa Riceputi \\ University of Bergamo, Italy\\ Istituto Nazionale di Fisica Nucleare - Sezione di Pavia, Italy}
\author{Gianluigi Zampa \\ Istituto Nazionale di Fisica Nucleare - Sezione di Trieste, Italy}

\abstract{The General Antiparticle Spectrometer (GAPS) is an Antarctic balloon-borne mission to
indirectly search for dark matter through sensitive observation of cosmic antiparticles. The first
flight is planned for late 2021. GAPS is the first experiment optimized specifically for detection of
low-energy (< 0.25 GeV/n) antideuterons, which are recognized as distinctive signals from dark
matter annihilation or decay in the Galactic halo. To achieve high sensitivity to cosmic antinuclei in
this low-energy range, GAPS uses a novel particle identification method based on exotic atom
capture and decay.

The GAPS instrument consists of ten planes of 1440 10 cm-diameter, 2.5 mm-thick, 8-strip lithium drifted silicon (Si(Li)) detectors, which constitutes the tracker, surrounded by a plastic scintillator time-of-flight system. A new fabrication technique has been developed to satisfy the stringent
requirements of the mission.

In this contribution, we describe the front-end electronics of the tracker of GAPS. The system is
composed of front-end ASICs and power supplies. The ASICs provide readout and digitization of
the signal (with an 11-bit ADC) in a wide dynamic range (10 keV - 100 MeV). Every ASIC has 32
channels and performs the readout for 4 detectors, for a total amount of 11520 channels. The
ASIC analog front-end is based on a dynamic compression technique to handle a large range of
signal amplitudes and features a low noise performance, achieving the required 4 keV resolution
at low energies. The power system supplies both bias voltages for the Si(Li) detectors and low
voltages for the electronics.}

\FullConference{36th International Cosmic Ray Conference -ICRC2019-\\
		July 24th - August 1st, 2019\\
		Madison, WI, U.S.A.}

\begin{document}

\section{Introduction}

Nowadays the nature of dark matter is one of the most relevant scientific problems for the understanding of the universe. The General Antiparticle Spectrometer (GAPS) aims to observe indirect signatures of dark matter through the identification of low-energy (<0.25 GeV/n) antideuterons in cosmic rays \cite{bib:gaps_1}, using an exotic atom technique. This technique and its unique event topology will give GAPS a nearly background-free detection capability that is critical in a rare-event search. GAPS is designed to carry out its science program using long-duration balloon flights in Antarctica. The first flight is scheduled in late 2021 \cite{bib:gaps_2}.

The instrument consists of a tracker composed of ten layers of lithium-drifted Silicon (Si(Li)) detectors, surrounded on all sides by a plastic scintillator time-of-flight (ToF) \cite{bib:gaps_3}. The ToF provides dE/dx energy loss and velocity measurement of the incoming ionizing particle as well as high-speed trigger and veto. The tracker acts both as target material for antiparticles annihilation and as a tracking device for incoming antinucleus and outgoing exotic atom products.

\subsection{The Si(Li) tracker}

The tracker is required to slow and capture antinuclei, measure 20 to 100 keV X-rays with FWHM <4 keV energy resolution and detect tracks of incident particles and exotic atom products.  In order to provide sufficient depth to stop incoming anti-nuclei (with kinetic energy up to 0.25 GeV/n) the array thickness must be greater than 25 mm having over $90 \% $ of the silicon thickness as its sensitive layer. To achieve sensitivity to the very low antideuteron flux,  more than 10 m$^{2}$ of active Si area is required. Since limited power is available to the cooling system on the balloon flight, the operation temperature is -35 to -45 C, significantly higher than that of typical silicon X-ray detectors \cite{bib:si_li_1}.

\begin{figure} 
	\centering
	\includegraphics[width=0.65	\linewidth]{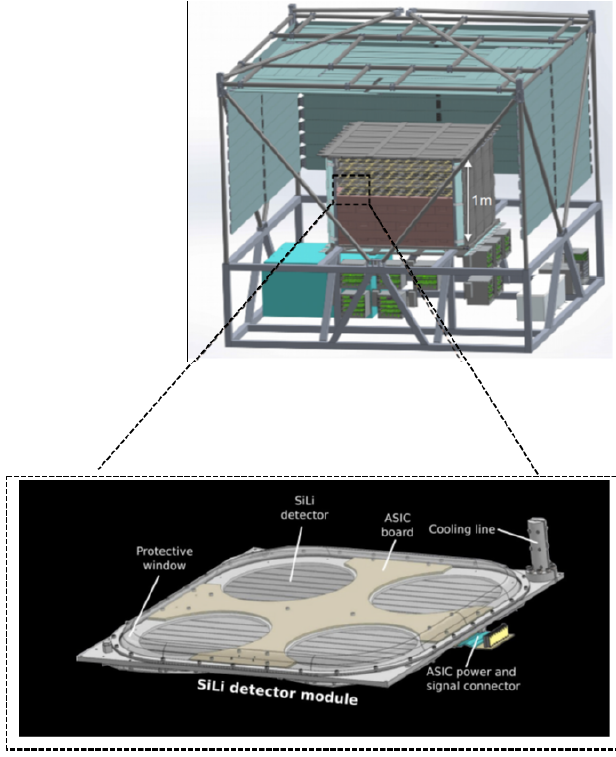}
	\caption{GAPS detector design. The TOF system surrounds a $1 m^{3}$ volume, filled with the silicon detector array nested into expanded polystyrene blocks. The insert shows an expanded view of a single Detector Module.}
	\label{fig:GAPS2}
\end{figure}

The Si(Li) tracker of the GAPS instrument is designed with a modular structure. The building block is comprised of 2$\times$2 matrix of detectors of 10 cm diameter and 2.5 mm thickness, segmented into 8 strips. Every matrix is mounted into an aluminum module, the detector module,  which provides the interface with cooling, power and read-out systems  (see the insert in Fig. \ref{fig:GAPS2}). The detector modules are nested into expanded polystyrene blocks. Those blocks supply insulation and protection during parachute shock and landing, at the same time, they have minimal interference with particle detection. The strips of each detector plane are all oriented in the same direction and are arranged orthogonally for alternate planes. The silicon detector array is ten layers deep and each tracking plane is composed of 12$\times$12 Si(Li) wafers \cite{bib:si_li_2}. 

The Si(Li) sensors are kept at the operational temperature of -43 $^{\circ}$C by means of an Oscillating Heat Pipe (OHP) passive cooling system, which has been custom developed for the GAPS experiment \cite{bib:Okazaki}.

The Si(Li) tracker electronics consist of three main subsystems: front-end electronics, back-end electronics and power system. Each detector module is read out by a custom 32 channel readout ASIC; a front-end board hosts the ASIC and its connection to the detector module. The back-end electronics board configures, controls and acquires data from the front-end boards. A chain of 6 detector modules (hence 6 ASICs) is controlled by one back-end channel through SPI protocol. Each tracker plane is controlled by one 6-channel FPGA-based DAQ box. The whole system comprises 10 DAQ boxes. The power system provides the bias voltage (HV) to the 360 detector modules of the Si(Li) tracker and the voltages (LV) to the front-end ASIC boards. The front-end electronics and the power system are described in detail in the next paragraphs. 

\section{The front-end electronics of the tracker}

The readout electronics must handle a large range of signal amplitudes (10 keV - 100 MeV) and provide low noise performance in order to achieve the required 4 keV resolution at low energies. The front-end board is placed in the center of the Si(Li) detector module (insert of Fig. \ref{fig:GAPS2}). It hosts the custom 32-channel ASIC, known as the Silicon LIthium DEtectors Readout (SLIDER32), which has been developed to read out the 11520 Si(Li) strips of the Silicon tracker and its connection to the detectors. The front-end board also forwards the bias voltage to the sensors, provides the bias voltage and the control signals to the ASIC and propagates those signals throughout the whole tracker.  

\subsection{The SLIDER32 ASIC}

\begin{figure} 
	\centering
	\includegraphics[width=0.9 	\linewidth]{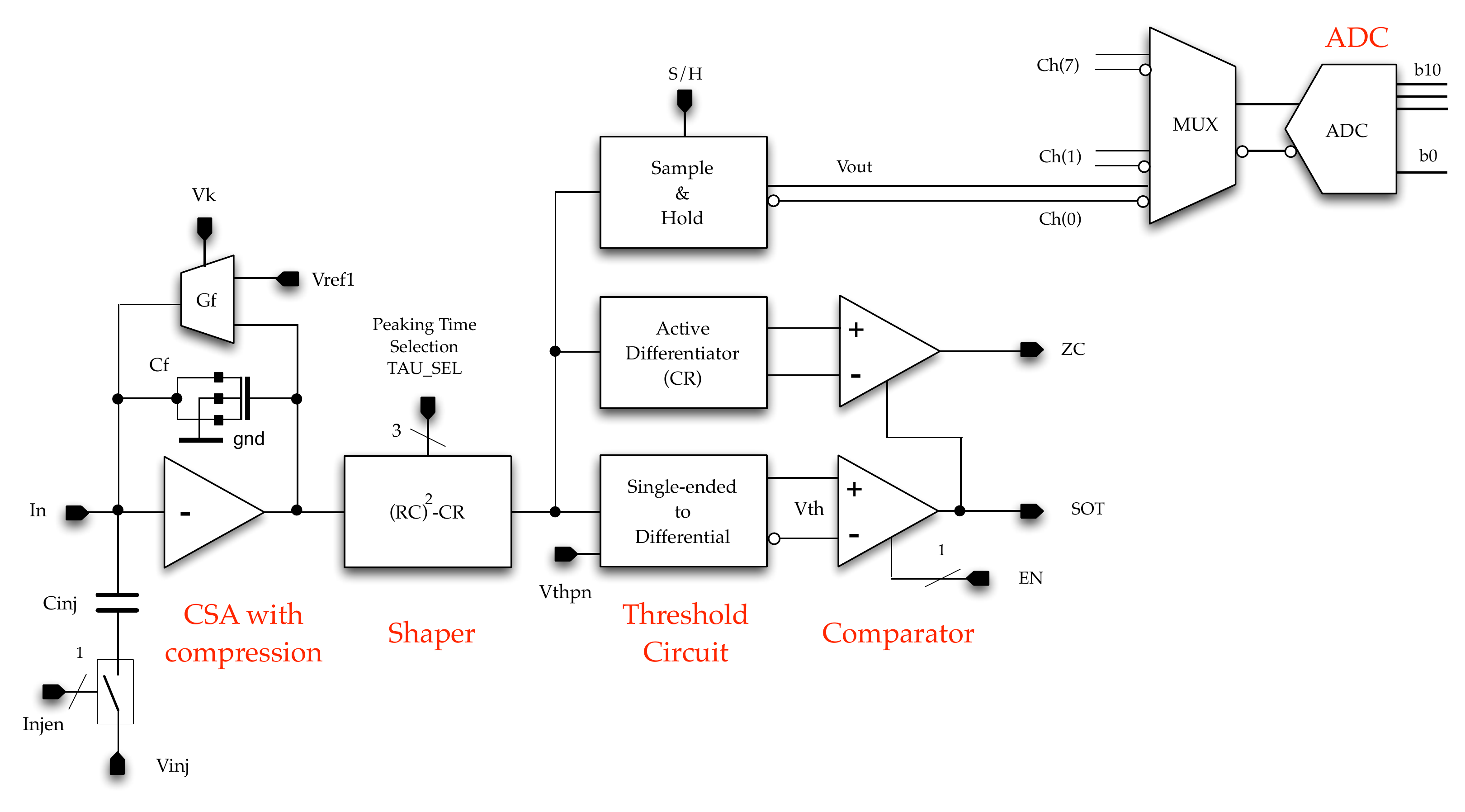}
	\caption{Simplified block diagram of the analog readout channel integrated in the 8-channel ASIC prototype. }
	\label{fig:ASIC2}
\end{figure}

The core of the SLIDER32 is the analog readout channel, whose block diagram is shown in Fig. \ref{fig:ASIC2}. In order to achieve the requirements on the wide range of signal amplitudes, the analog conditioning scheme is based on a low-noise charge-sensitive amplifier (CSA) featuring dynamic signal compression \cite{bib:asic}. This solution takes advantage of the non-linear features of a MOS capacitor in the feedback loop of the charge-sensitive preamplifier itself. In the CSA, a continuous reset is provided by a Krummenacher feedback network. 

The amplifier is followed by a unipolar second-order semi-Gaussian filter. 
%The shaping actually takes place in two steps. First the signal from the preamplifier undergoes an integration, then the shaping is completed by an active filter which provides one more integration and a differentiation and also implements a gain function. 
The signal peaking time at the filter output can be selected among eight values (from 250 ns up to 2 $\mu$s). 

After the filter, the analog conditioning scheme is split into three different paths. On one side, the signal is converted from single-ended to differential and then it is compared to a preset differential threshold of a discriminator.
%a threshold circuit converts the single-ended signal at the shaper output to a differential one and a differential DC threshold voltage is superimposed to the dynamic signal to drive the comparator.
On the second path, a single-ended to differential Sample\&Hold provides a signal proportional to the shaping peak to the subsequent differential SAR (Successive Approximation Register) ADC for analog information detection. On the third branch, the filtered signal is differentiated. The identification of the zero-crossing of the resulting bipolar signal provides a trigger, synchronous with the shaper peaking time, for the single-ended to differential Sample\&Hold.

The first step towards the final ASIC has been the design of two prototypes named SLIDER4 and SLIDER8. The characterization activity is now in progress and will continue in the coming months. The first prototype of the final ASIC, which will host 32 channels and an 11-bit ADC, has been submitted in Spring 2019.

\subsection{The power system}

The power system is composed of two subsystems: the High Voltage Power Supply (HVPS) system  to supply HV to the detector modules, and the Low Voltage Power Supply (LVPS) system to provide LV to the front-end boards. 

The bias voltage of the Si(Li) detectors varies in the range 150--300\,V, and the supplied voltage has to be provided with an accuracy of 1V. The typical absorbed current is 0.1 nA and must be monitored by the HVPS with a precision of 1-2 nA.  Other driving requirements for the design of the HVPS are the possibility to reach the desired bias voltage through several (programmable) steps (ramp-up) and the environment requirements (wide temperature operative range, low atmospheric pressure, ...).

The front-end ASIC board needs four different voltages. In the power distribution scheme adopted,  one LV channel provides voltage to a chain of six detector modules, for an estimated total power of 15W. The LV has to meet the requirements on the maximum power of each channel, the total number of channels (60) and the environment, while providing a voltage and current monitor resolution respectively of 1 mV and 0.7 mA. 

In addition, it is necessary to consider that the Si (Li) detectors cannot be brought to the working voltage before the front-end electronics have been powered, and a similar procedure must be followed even when switching off the bias voltage. It is therefore necessary to equip the power supply system with an inter-lock logic that prevents the occurrence of incorrect switching on or off sequences that could damage the detectors. 

\begin{figure} 
	\centering
	\includegraphics[width=0.9 	\linewidth]{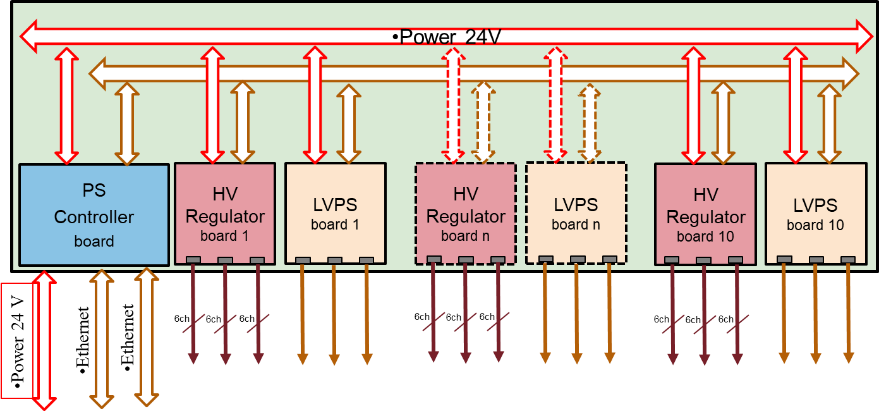}
	\caption{A schematic diagram of one of the power system hosted in the Eurocard sub-rack. Each board can control 18 channels. }
	\label{fig:power}
\end{figure}

A custom power system able to manage the 360 HV channels (one per each Si(Li) detector module) and the 60 LV channels (one per chain of six Si(Li) detector modules) has been designed. A scheme of the power distribution scheme is presented in Fig. \ref{fig:power}. The power system is based on pairs of HVPS and LVPS boards hosted on double Eurocard boards. Each pair is able to supply all the voltages necessary for the operation of a half-plane of the tracker. %Dieci coppie di schede forniscono le tensioni necessarie ad alimentare i cinque piani di tracciatore con le strip disposte lungo la medesima direzione e altrettante schede sono necessarie per i cinque piani con le strip orientate in direzione ortogonale. 

The power system is equipped with an Ethernet interface in order to properly handle the power on and off, voltages and currents set-up sequences, and the monitoring of telemetry data. The redundant Ethernet interface is performed by a controller board based on a MicroController. The MicroController is the main interface of the system with the DAQ system (via Ethernet protocol) and provides control of the HVPS and LVPS boards hosted in the sub-rack. 
%I collegamenti tra il controllore e le schede presenti nel cestello, nonché tutte le connessioni tra le singole schede (segnali di inter-lock) sono realizzate attraverso il backplane presente sul cestello. Sul backplane viaggiano anche i 24V del bus primario e le tensioni di alimentazione secondarie. 
%Le schede sia HV che LV sono state progettate con formato di forma doppio Eurocard. Quindi due cestelli Eurocard da 6U sono in grado di ospitare l’intero sistema Lo schema a blocchi di figura 2 mostra l’organizzazione delle schede nel cestello Eurocard.

A prototype of the power system has been designed and fabricated. The testing of the prototype is under way.  

\section{Conclusions}

In this paper, we described the front-end electronics of the Si(Li) detector of the GAPS experiment. The system is composed of readout ASICs and power supplies. The testing of the prototypes is under way. The development of the front-end electronics will be completed by early summer 2020, on schedule for the launch foreseen in late 2021.

\section*{Acknowledgments}

This work is supported in the U.S. by NASA APRA grants (NNX17AB44G, NNX17AB45G, NNX17AB46G, and NNX17AB47G), in Japan by JAXA/ISAS Small Science Program and JSPS KAKENHI grants (JP26707015 and JP17H01136), in Italy by INFN and the Italian Space Agency through the GAPS ASI INFN agreement n. 2018-28-HH.0.

\end{document}